\documentclass[aps,prb,reprint,superscriptaddress,amsfonts,amssymb,amsmath]{revtex4-2}

\usepackage{graphicx}
\usepackage{dcolumn}
\usepackage{bm}
\usepackage[mathlines]{lineno}
\usepackage{siunitx}
\usepackage{booktabs}
\usepackage{color}
\usepackage{multirow}

\begin{document}


\title{NMR determination of van Hove singularity and Lifshitz transitions in nodal-line semimetal ZrSiTe}

\author{Yefan Tian}
\email{yefantian93@gmail.com}
\altaffiliation{Present address: LASSP \& Department of Physics, Cornell University, Ithaca, NY 14853, USA}
\affiliation{Department of Physics and Astronomy, Texas A\&M University, College Station, TX 77843, USA}
\author{Yanglin Zhu}
\affiliation{Department of Physics, Pennsylvania State University, University Park, PA 16802, USA}
\author{Rui Li}
\affiliation{Department of Physics and Astronomy, Texas A\&M University, College Station, TX 77843, USA}
\author{Zhiqiang Mao}
\affiliation{Department of Physics, Pennsylvania State University, University Park, PA 16802, USA}
\author{Joseph H. Ross, Jr.}
\email{jhross@tamu.edu} 
\affiliation{Department of Physics and Astronomy, Texas A\&M University, College Station, TX 77843, USA}

\date{\today}

\begin{abstract}
We have applied nuclear magnetic resonance spectroscopy to study the distinctive network of nodal lines in the Dirac semimetal ZrSiTe. The low-$T$ behavior is dominated by a symmetry-protected nodal line, with NMR providing a sensitive probe of the diamagnetic response of the associated carriers. A sharp low-$T$ minimum in NMR shift and $(T_1T)^{-1}$ provides a quantitative measure of the dispersionless, quasi-2D behavior of this nodal line. We also identify a van Hove singularity closely connected to this nodal line, and an associated $T$-induced Lifshitz transition. A disconnect in the NMR shift and line width at this temperature indicates the change in electronic behavior associated with this topological change. These features have an orientation-dependent behavior indicating a field-dependent scaling of the associated band energies.
\end{abstract}

\maketitle


The recently discovered nodal-line semimetals (NLSMs) have been widely investigated due to distinctive topological features including a high density of bulk topological states and drumhead surface states. These present intriguing possibilities for applications \cite{burkov2011topological,fang2015topological,lv2016extremely,sankar2017crystal,hu2017nearly}. Among NLSMs, the Zr$XY$ system ($X$ = Si, Ge, Sn; $Y$ = S, Se, Te) is the most studied, because of their nodal-line network close to the Fermi energy \cite{klemenz2019topological}. This can lead to enhanced correlation effects, recently indicated to be important in ZrSiSe and ZriSiS \cite{rudenko2018excitonic,shao2020electronic,gatti2020light} and associated with an unconventional mass enhancement and predicted semimetal-exciton insulator transition \cite{pezzini2018unconventional,wang2020quantum}.

ZrSiTe is distinct among this family since it exhibits a non-symmorphic symmetry-protected nodal line very close to the Fermi energy \cite{topp2016non}. Its large region of topologically protected drumhead surface states \cite{muechler2020modular} as well as very large spin Berry curvature and associated spin Hall and Nernst response \cite{yen2020tunable} suggest spintronic and related applications. Suggested changes of Fermi surface topology in this family include a temperature-induced Lifshitz transition in ZrSiSe \cite{chen2020temperature}, and a pressure-induced Lifshitz transition in ZrSiTe \cite{krottenmuller2020indications} as well as indications of photo-induced phonon driven transformation \cite{kirby2020signature}.

In this Letter, we have utilized nuclear magnetic resonance (NMR) spectroscopy as a highly sensitive probe of states close to the Fermi energy. We show that temperature-induced changes lead the chemical potential to coincide with the protected node at low temperatures. In addition, based on density functional theory (DFT) calculations, we demonstrate that the arrangement of parallel Dirac nodes leads to a 2D van Hove singularity (VHS) near $E_F$. This leads to temperature-induced changes, including a temperature-induced Lifshitz transition.

\begin{figure}
\includegraphics[width=0.8\columnwidth]{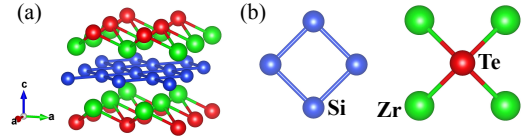}%
\caption{\label{structure} (a) Tetragonal crystal structure of ZrSiTe (space group: $P4/nmm$). (b) Square-net structure of Si layer and Zr-Te layer.}
\end{figure}


\begin{figure*}
\includegraphics[width=2\columnwidth]{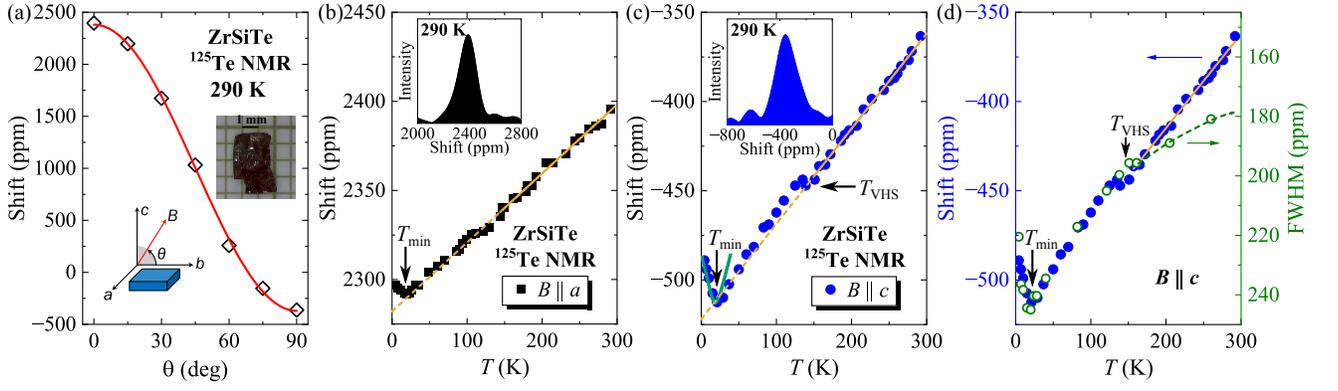}%
\caption{\label{shift} (a) Angular dependence of ZrSiTe NMR with a fit to $K=K_\mathrm{iso}+\Delta K\cdot(3\cos^2\theta-1)/2$. Inset: representative crystal from growth batch. Shift vs temperature for (b) $B\parallel a$ and (c) $B\parallel c$, with spectra inset. Linear fits at high $T$ shown for both cases. Low-$T$ solid curve in (c): orbital diamagnetism model. (d) $B \parallel c$ line width (open circles) plotted with the shift (solid circles). Curves: guides to the eye.}
\end{figure*}

For the experiments, the ZrSiTe single crystals were prepared using chemical vapor transport method. The stoichiometric mixture of Zr, Si and Te powder was sealed in a quartz tube with iodine being used as transport agent (2 mg/cm$^3$). Plate-like single crystals with metallic luster can be obtained via the vapor transport growth with a temperature gradient from 950\,$^\circ$C to 850\,$^\circ$C. NMR experiments utilized a custom spectrometer at fixed field $B=9$ T, with a sample of six crystals from the same batch assembled edge to edge with axes aligned, and with the field parallel to $c$ ($B\parallel c$) and $a$ ($B\parallel a$). $^{125}$Te relative frequency shifts [$K=(f-f_0)/f_0$] were calibrated by aqueous Te(OH)$_6$ and adjusted for its 707 ppm paramagnetic shift to the dimethyltelluride standard ($f_0$) \cite{inamo1996125te}. 

DFT calculations were also carried out using the PBE potential \cite{perdew1996generalized} as implemented in the \textsc{WIEN2k} code \cite{blaha2020wien2k}. Experimental structure parameters $a=3.71$ \si{\angstrom}, $c=9.51$ \si{\angstrom}, with Zr $z = 0.2238$ and Te $z = 0.6399$, were used \cite{bensch1994structure,hu2016evidence,krottenmuller2020indications} with cutoff parameter $k_\mathrm{max} = 7/R_\mathrm{MT}$ inside the interstitial region. A mesh of 3000 $k$-points was employed for initial calculations. Fermi surfaces and densities of states were also calculated on a mesh of 10000 irreducible $k$-points and rendered using \textsc{XCrySDen} \cite{kokalj1999xcrysden}.



$^{125}$Te NMR spectra [insets, Fig.~\ref{shift}(b) and \ref{shift}(c)] exhibit a peak corresponding to the single Te site in ZrSiTe. The angular dependence of the shifts are shown in Fig.~\ref{shift}(a) with $\theta$ defined in the inset. The results were fitted [red curve in Fig.~\ref{shift}(a)] to $K=K_\mathrm{iso}+\Delta K\cdot(3\cos^2\theta-1)/2$, where $K_\mathrm{iso}=547\pm14$ ppm is the isotropic shift and $\Delta K=1834\pm23$ ppm.

The temperature-dependent shifts for $B\parallel a$ (denoted $K_{\parallel a}$) and $B\parallel c$ ($K_{\parallel c}$) are shown in Figs.~\ref{shift}(b) and \ref{shift}(c). These exhibit V-shaped minima at $T_\mathrm{min}= 20$ K, with a much larger slope for $B\parallel c$. With ZrSiTe exhibiting a one-dimensional protected Dirac nodal line very close to the Fermi energy \cite{topp2016non}, the large shift changes for $K_{\parallel c}$ can be attributed to the diamagnetism associated with this nodal line. At higher temperatures $K$ for both orientations continues to increase, with the exception of the small drop in $K_{\parallel c}$ at 140 K.

As shown in Fig.~\ref{shift}(d), the $B\parallel c$ full width at half maximum (FWHM) also exhibits a V-shaped anomaly with a maximum at 20 K (with the FWHM scale inverted to properly match the shift). For $B\parallel a$ the width is smaller and nearly constant. The $B\parallel c$ widths scale with $K_{\parallel c}$ below 140 K as shown in Fig.~\ref{shift}(d), but the change in slope above 140 K indicates a change of broadening mechanism at high $T$. The width is scaled to $-0.7$ times the shift in the figure, consistent with an origin for the low-$T$ broadening due to a distribution of local fields associated with the diagnetism of the plate-shaped crystals, with the maximum occurring where the V-shaped diamagnetic shift has the largest magnitude.

\begin{figure*}
\includegraphics[width=1.95\columnwidth]{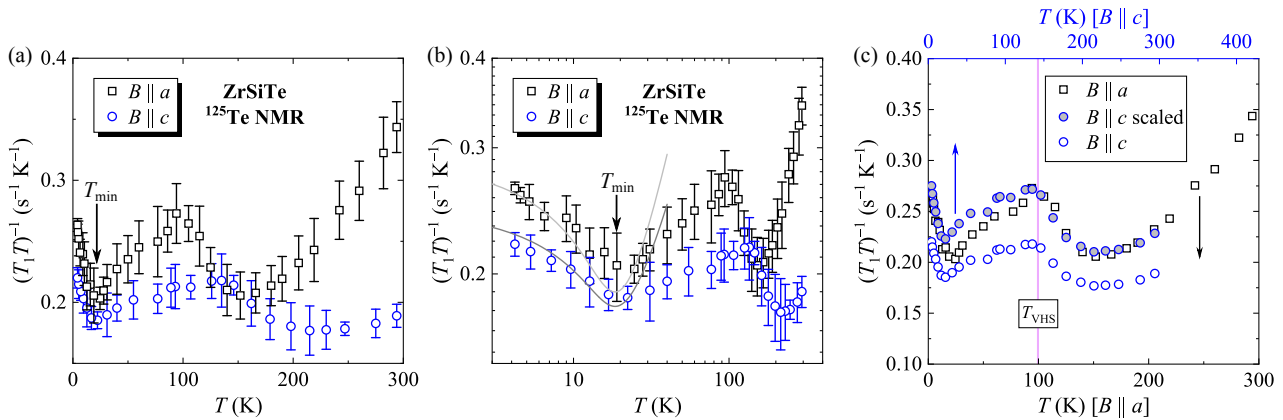}%
\caption{\label{T1} $(T_1T)^{-1}$ vs $T$ for $B\parallel a$ and $B\parallel c$ in (a) linear and (b) log scale. Curves show orbital-mechanism calculation based on protected Dirac nodal line. (c) Comparison between scaled $(T_1T)^{-1}$ values for $B \parallel c$ and unscaled $B \parallel a$ results.}
\end{figure*}


Spin-lattice relaxation measured by inversion recovery was well-fitted to a single exponential $M(t)=(1-Ce^{-t/T_1})M(\infty)$, giving $(T_1T)^{-1}$ values shown in Figs.~\ref{T1}(a) and \ref{T1}(b). While in simple metals $(T_1T)^{-1}$ is constant, we find sharp $(T_1T)^{-1}$ minima at $T_\mathrm{min}=20$ K for both orientations, with an approximately linear increase on both sides. These coincide with the observed shift minima. 

At higher temperatures, there is a large orientation-dependence. $(T_1T)^{-1}_{\parallel a}$ reaches a maximum at 100 K, and goes through a minimum near 150 K, while for $B\parallel c$, these are shifted to 140 K and 225 K, respectively. The maximum at 140 K coincides with the feature in $K_{\parallel c}$ and the break in the line width scaling [Fig.~\ref{shift}(d)], which collectively indicates a change in electronic behavior at this temperature.

\begin{figure*}
\includegraphics[width=1.85\columnwidth]{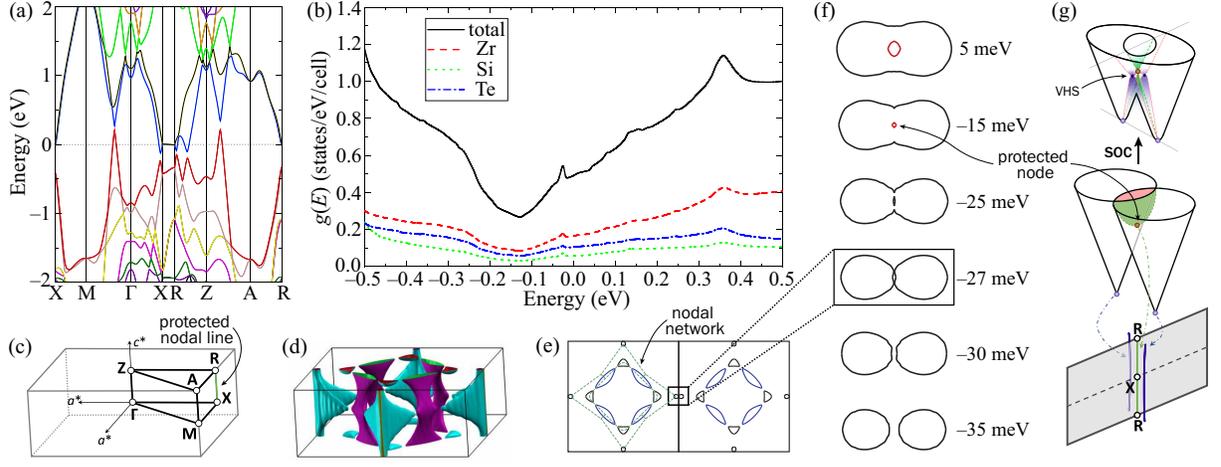}%
\caption{\label{dft} (a) ZrSiTe band structure with spin-orbit coupling. (b) Density of states near $E_F$ (defined as $E=0$). (c) Tetragonal Brillouin zone with high-symmetry points. (d) Fermi surface with hole surfaces purple (darker) and electron surfaces green. (e) Fermi surface slice in Z-A-R plane with sketched nodal network. (f) Slices near R at energies shown. (g) Sketch of parallel nodal lines near R and X, with intersecting Dirac cones and effects of spin-orbit coupling illustrated.}
\end{figure*}

The band structure is shown in Fig.~\ref{dft}(a). In agreement with previous reports \cite{topp2016non}, a gapless Dirac nodal line extends from R to X [$c^*$ direction, Fig.~\ref{dft}(c)], very close to $E_F$, and protected against spin-orbit induced gap opening by non-symmorphic symmetry. From the band dispersions, we calculate mean Fermi velocities ($v_F$) $5.6\times10^5$ m/s at X, and $3.1\times10^5$ m/s at R. We use the average value, $v_F = 4.3\times10^5$ m/s, in the analysis below. 

In addition to the protected nodal line, there is a cage-like network of Dirac nodal lines \cite{schoop2016dirac,hu2016evidence} with a finite gap due to spin-orbit coupling. The nodal network includes segments enclosed by the large electron and hole-like Fermi surfaces in Fig.~\ref{dft}(d), connected by loops in the $\Gamma$-X-M and Z-R-A planes as sketched in Fig.~\ref{dft}(e) (modeled after \cite{muechler2020modular}). The density of states $g(E)$ [Fig.~\ref{dft}(b)] is dominated near $E_F$ by this nodal network, where similar to ZrSiS the excitations are essentially all Dirac fermions \cite{fu2019dirac}.

The protected nodal line runs parallel to segments of the nodal network, as sketched in Fig.~\ref{dft}(g). Constant-energy slices from the Z-R-A plane [Figs.~\ref{dft}(e) and \ref{dft}(f)] further illustrate this: 35 meV below $E_F$ two disconnected regions enclose only the nodal-network segments, while for $-25$ meV and above these regions have merged, with the protected nodal line enclosed in the disconnected region in the center. Similar features can be seen in the $\Gamma$-X-M plane. A topology change occurs at $-27$ meV where the inner and outer regions come into contact along R-A. 

The contact points at $-27$ meV are saddle points, representing local minima along R-A and transverse maxima. To a good approximation, because of the nearly vanishing dispersion along R-X, this is a 2D van Hove singularity (VHS), with an associated log energy dependence. The corresponding peak in $g(E)$ is seen in Fig.~\ref{dft}(b). Such features near $E_F$ have been proposed to be important for superconductivity in bulk KFe$_2$As$_2$ \cite{fang2015observation} and PdTe$_2$ \cite{kim2018importance}. Here, the origin of the VHS is similar to twisted bilayer graphene, where an avoided crossing between closely-spaced Dirac nodes leads to a saddle point contributing to its anomalous electronic behavior \cite{luican2011single,brihuega2012unraveling,andrei2020graphene}. In the present case, symmetry forbids the avoided crossing along the R-X centerline producing the protected nodal line.



Regarding the $K$ and $(T_1T)^{-1}$ minima near 20 K, the well-known Korringa relation \cite{carter1977metallic} can connect $K$ and $(T_1T)^{-1}$, such as occurs due to electron spins in simple metals. This relation is $K^2T_1T=\hbar \gamma_e^2/(4\pi k_B\gamma_n^2) = 2.6\times10^{-6}$ s\,K, where $\gamma_{e(n)}$ is the electron (nuclear) gyromagnetic ratio, and the numerical value is for $^{125}$Te. $K_{\parallel c}$ drops by 30 ppm from 4 K to 20 K while $(T_1T)^{-1}_{\parallel c}$ is reduced by 0.03 s$^{-1}$\,K$^{-1}$ [Figs.~\ref{shift}(c) and \ref{T1}(b)], giving $K^2 T_1 T=3 \times 10^{-8}$ s\,K, much smaller than expected due to the disappearance of a spin-hyperfine contribution. Similarly, for $B\parallel a$ the changes in $(T_1T)^{-1}$ are much larger than expected can be explained by spin mechanism, given the magnitude of $K_{\parallel a}$.

There is an extended-orbital mechanism for Dirac carriers that instead provides a good explanation for the large $(T_1T)^{-1}$ \cite{okvatovity2019nuclear,maebashi2019nuclear}. This mechanism includes a significant contribution from carriers far from the nucleus \textcolor{black}{without relying on local hyperfine fields}. Evidence for this mechanism has been seen recently in other zirconium tellurides \cite{tian2019dirac,tian2020topological}. For a quasi-2D nodal line, one obtains $(T_1T)_{\parallel a}^{-1} = (3/2) (T_1T)_{\parallel c}^{-1}$ \cite{maebashi2019nuclear}, close to what is observed at low $T$ (Fig.~\ref{T1}). The corresponding orbital mechanism for $K$ connects to bulk diamagnetism of the Dirac electrons; $K_{\parallel c}$ scales with the line width, helping to confirm that the bulk diamagnetism dominates. These mechanisms are not expected to follow a Korringa relation \cite{okvatovity2019nuclear}; however, the V-shaped minima in both the shift and $(T_1T)^{-1}$ can be explained on the basis of orbital coupling to the protected nodal line, as shown below.

We first consider the diamagnetic shift. The protected node will respond in much the same way as the 2D Dirac system in graphene, assuming the node is a linear two-band crossing with no dispersion along $c$. With the 9 T NMR field applied along $c$ and $v_F = 4.3 \times 10^5$ m/s, the Landau level energies are $\varepsilon_\mathrm{LL}=\pm\sqrt{2e\hbar v_F^2 B|N|}=\pm 47\,[\mathrm{meV}]\,\sqrt{|N|}$, neglecting Zeeman splitting. Each level holds a density $n_\mathrm{LL}=4B/(c \Phi_0)=9.2 \times 10^{18}$ cm$^{-3}$ \cite{tian2020topological}, with $\Phi_0=4.1 \times 10^{-15}$ T\,m$^2$ the flux quantum, $c=9.51$ \si{\angstrom}, and 4 representing the level degeneracy (2 for spin, 2 for the R-X valley degeneracy). As described in Ref.~\cite{li2015field}, the magnetization per volume is $M=-\partial(\Omega/V)/\partial B$, with the grand potential density $\Omega/V=-k_BTn_\mathrm{LL}\sum_N\ln[1+e^{(\varepsilon_\mathrm{LL}-\mu)/k_BT}]$, where $\mu$ is the chemical potential. \textcolor{black}{With all occupied levels in the sum, it can be seen that this differs from the Landau diamagnetism, which depends on carriers only at $E_F$.} This yields an NMR shift contribution $K_\mathrm{dia}=\mu_0 M/B$, the same as the SI dimensionless susceptibility to the extent that $M/B=\partial M/\partial B$, although in large fields $K$ becomes linear in $\mu$ and $T$ rather than exhibiting a $\mu=T=0$ singularity as found in the low field limit \cite{raoux2015orbital}. \textcolor{black}{The result is also distinct from the log singularity obtained in the 3D case \cite{maebashi2019nuclear}.}  For the present case, demagnetizing effects need also to be included. With crystal dimensions ($a \times b \times c$) close to $2 \times 2 \times 0.5$ mm$^3$, the mean demagnetizing factor from a cuboid estimation \cite{prozorov2018effective} is $N \approx 4ab/[4ab + 3c(a + b)] = 0.7$, giving an observed $K$ that is scaled by a factor of about 0.3. 

\begin{figure*}
\includegraphics[width=2\columnwidth]{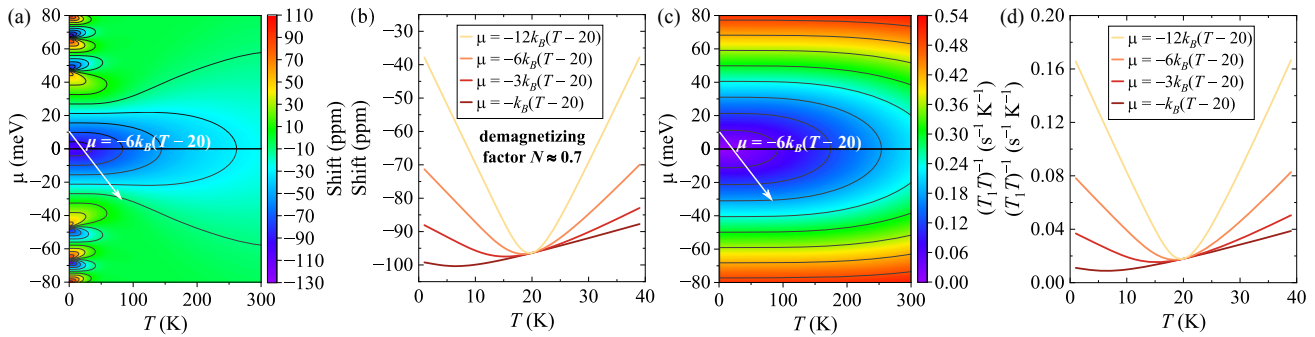}%
\caption{\label{discussion} (a) Heat map showing ($\mu, T$)-dependent $K_{\parallel c}$ calculated from nodal-line diamagnetism. (b) $K_{\parallel c}$ vs $T$ assuming $\mu(T)$ as shown. (c) Heat map showing calculated ($\mu, T$)-dependent $(T_1T)_{\parallel c}^{-1}$. (d) $(T_1T)_{\parallel c}^{-1}$ vs $T$ assuming $\mu(T)$ as shown.}
\end{figure*}

The resulting shift is displayed as a heat map in Fig.~\ref{discussion}(a), with the line-node energy defined as $E = 0$. We find that the V-shaped minimum (and corresponding line width maximum) can be obtained only if $\mu(T)$ crosses the node at $T_\mathrm{min}$. Assuming $\mu=-Ak_B (T-20\,[\mathrm{K}])$, Fig.~\ref{discussion}(c) shows curves calculated for various multipliers $A$. As expected \cite{li2015field}, there is a linear change in $K_\mathrm{dia}$ away from the node crossing as long as $\mu$ is between the $N = \pm1$ levels. Larger values of $A$ produce a relatively sharp minimum, similar to the measured results.

In Fig.~\ref{shift}(c), this model is compared to the data using $A = 6$, with good agreement below the 20 K node crossing. $\mu(T)$ in this case is shown by the white arrow in Fig.~\ref{discussion}(a). Together with the line width results, this is a good indication that indeed the bulk Dirac diamagnetism is the dominant contribution to the low-$T$ shift. There are previous reports \cite{wang2020landau,tay2020unusual,suetsugu2021giant} of diamagnetic shifts in point-node Dirac/Weyl systems;  here we provide a quantitative model for this behavior based on the quasi-2D protected node in ZrSiTe. Above 20 K, the data fall below the calculated curve, likely caused by a change in Landau level configuration as $\mu$ approaches the VHS [Fig.~\ref{dft}(f)].

Ref.~\cite{maebashi2019nuclear} gives  $(T_1T)^{-1}$ for the quasi-2D extended-orbital case. With zero Dirac gap for the protected nodal line, the result is
\begin{multline} \label{sl}
\bigg(\frac{1}{T_1T}\bigg)_{\parallel a}=\frac{3}{2}\bigg(\frac{1}{T_1T}\bigg)_{\parallel c}= \\
4\beta\frac{\mu_0^2\gamma_n^2e^2k_B}{(4\pi)^2} \int dE\bigg[-\frac{\partial f(E)}{\partial E}\bigg]\frac{E}{\hbar^2cv_F}\ln\frac{2|E|}{\hbar\omega_0},
\end{multline}
with $E=\pm v_Fk \hbar$, $c$ the lattice constant, and $f(E)$ the $\mu$- and $T$-dependent Fermi function. $\beta$ is a multiplier relating to the overall density of Dirac carriers within the unit cell \cite{okvatovity2019nuclear}, while the leading factor of 4 is due to the valley degeneracy since the Fermi golden rule mechanism introduces a product of source and final state multiplicities.

Fig.~\ref{discussion}(c) shows a heat map for $(T_1T)_{\parallel c}^{-1}$ computed with $c=9.51$ \si{\angstrom}, $v_F=4.3\times10^5$ m/s and $\beta=1.2\times 10^{-4}$. The white arrow represents $\mu=-Ak_B(T-20\,[\mathrm{K}])$, with $A = 6$ from the shift analysis, while Fig.~\ref{discussion}(b) shows results for other multipliers. The linear behavior away from the node is characteristic of the quasi-2D case \cite{maebashi2019nuclear}, with a point node instead exhibiting quadratic behavior. Larger values of $A$ generate the relatively sharp minimum seen in the data. The curves in Fig.~\ref{T1}(b) were generated from this model using $A=6$, including the factor $3/2$ for $B \parallel c$. Both curves include an additive constant 0.175 (s\,K)$^{-1}$, due to nodal-network carriers at $E_F$ not belonging to the protected node, assumed to have little temperature dependence at these temperatures. While the agreement is particularly good below 20 K, the curves fall above the data for $T > 20$ K. Similar to the shift results, this is likely due to the VHS.

While the shift analysis has only the overall multiplier $A$ as adjustable parameter, for $(T_1T)^{-1}$ the orbital mechanism also includes the multiplier $\beta$. It is difficult to know this parameter \textit{a priori}; however, comparison shows that the results are in the expected range: ZrTe$_5$ with its point-node exhibits a nearly-quadratic $T$- and $\mu$-dependent $(T_1T)^{-1}$ \cite{tian2021gap} fitted to $(T_1T)^{-1} \cong 40\mu^2$ (s\,K\,eV$^2$)$^{-1}$, or a 0.004 (s\,K)$^{-1}$ difference as $\mu$ increases from zero to 10 meV. This is the same as the 20 K to 4 K change in $\mu$ modeled here [Fig.~\ref{discussion}(a)], for which $(T_1T)_{\parallel a}^{-1}$ changes by 0.055 (s\,K)$^{-1}$, 14 times larger than for ZrTe$_5$ for the same $\Delta \mu$. The valley degeneracy plus the additional phase space for electron scattering in the line node case leads to \cite{tian2020topological} a ratio $(T_1T)_{\parallel a}^{-1}/(T_1 T)_\mathrm{point}^{-1}=\frac{3v_F}{2\pi\hbar\mu c}$ for the line node vs the point-node, where the small differences in $v_F$ between materials has been neglected. This ratio is 13, consistent with the scaled comparison of relaxation rates. 

The model in which the protected line is a quasi-2D nodal line with very little dispersion agrees well with the data, for example even 10 meV dispersion along this line will lead to considerable broadening of the V-shaped shift minima, which is not observed. Also the factor 3/2 in Eq.~(\ref{sl}) requires $\Delta k_z \neq 0$ scattering among degenerate states along the line. These factors together suggest that the protected nodal line is indeed very close to the quasi-2D case. 

For $B\parallel a$, there will be no Landau quantization and thus the Dirac diamagnetism vanishes. The $K_{\parallel a}$ minimum must therefore be due to a spin contribution which presumably also underlies the larger $K_{\parallel c}$ minimum. The spin shift is proportional to $g(E_F)$ \cite{carter1977metallic}. Since $g(E_F) \propto |E|$ for the quasi-2D case, this will also give a V-shaped shift as $\mu$ crosses the node energy. However, the spin $(T_1T)^{-1}$ contribution is quadratic and as noted above based on the Korringa product, this term will be negligible compared to the observed $(T_1T)^{-1}$ magnitude.


The sharp VHS adjacent to the protected nodal line implies that additional electronic features relying upon the enhanced $g(E)$ could be engineered into this system \cite{fang2015observation,kim2018importance,luican2011single,brihuega2012unraveling,andrei2020graphene}, by tuning $E_F$ to the VHS. Tuning could be accomplished by chemical pressure \cite{hosen2017tunability}, physical pressure or strain \cite{krottenmuller2020indications,zhou2020effect}, and light activation \cite{kirby2020signature} with the required energy difference being quite small. This could be another avenue for switching the topological behavior for quantum computing applications. Note that a VHS may also be present due to extrema of the nodal network \cite{shao2020electronic}; however, such 3D saddle points would not give the sharp features we observe.

At 140 K, there is a disconnect for $B \parallel c$, seen in the shift vs $T$ [Fig.~\ref{shift}(c)] as well as the line width scaling [Fig.~\ref{shift}(d)]. Based on the model described above, we infer that this occurs when $\mu$ crosses the VHS as it continues to decrease relative to the line-node energies, thus this change corresponds to a $T$-induced Lifshitz transition. We label this transition temperature $T_\mathrm{VHS}$. There is also a broad maximum in $(T_1T)^{-1}$ at this temperature, although the maximum occurs at a reduced temperature of about 100 K for $B \parallel a$. 

To show that the behaviors for the two orientations are closely related, Fig.~\ref{T1}(c) shows $(T_1T)^{-1}$ with the temperature scale reduced by a factor of 0.7 for $B \parallel c$. For the additional scaled $B \parallel c$ plot, a common baseline of 0.11 (s K)$^{-1}$ was subtracted, and then the remainder was scaled by the $3/2$ factor appearing in Eq.~(\ref{sl}). The broad minimum at temperatures above $T_\mathrm{VHS}$ for $B \parallel c$ follows the same trend as for the $B \parallel a$ data, indicating that only the relative energy scale is modified by the field. 

Examination of the band-structure shows that the obvious $g(E)$ minimum near $-130$ meV [Fig.~\ref{dft}(b)] is associated with the nodal lines parallel to R-X that contribute to the VHS, so it is reasonable to associate the $(T_1T)^{-1}$ minima above $T_\mathrm{VHS}$ with $\mu$ crossing this feature. However, the roughly quadratic shape of the minimum contrasts with the V-shape for the crossing near 20 K. This points to a more conventional local-orbital mechanism for the high-$T$ $(T_1T)^{-1}$, since \cite{slichter1990principles} normally $(T_1T)^{-1} \propto \int d\varepsilon \frac{df}{d\varepsilon}g^2(\varepsilon)$, and with $g(\varepsilon) \propto \varepsilon$ for the line node, this will give $(T_1T)^{-1} \propto \mu^2$.

The relatively large change of $\mu$ at low $T$ cannot be explained on the basis of charge conservation alone. Indeed, from the calculated $g(E)$ at $E_F$ (Fig.~\ref{dft}), standard considerations \cite{ashcroft1976solid} lead to a change of $\mu(T)$ by only a few meV up to room temperature within a rigid-band picture. On the other hand, based on the consistent picture provided by the shifts, the $(T_1T)^{-1}$, and line width it is clear that $\mu(T)$ has a large $T$-dependence and crosses the protected node at low $T$. In ZrSiSe, transport results were recently proposed to give evidence for even larger $T$-induced changes of $\mu$ \cite{chen2020temperature}. For the present case, we assume it is relative shifts of band energies that contribute to this process causing the node to shift relative to $\mu$, perhaps induced by the promotion of carriers from donor states, in a manner akin to the changes that have been proposed \cite{kirby2020signature} to accompany the optical promotion of carriers in ZrSiTe.

With the $T_1$ minima at high $T$ understood as due to $\mu$ crossing the vertical line-node parts of the nodal network, the absence of a corresponding feature in $K$ can be understood in light of the relatively large gap in the nodal-network system. With a gap, the diamagnetic shift is greatly reduced, even though the orbital mechanism can still induce a significant $(T_1T)^{-1}$ contribution \cite{maebashi2019nuclear}. The origin of the apparent temperature scaling above $T_\mathrm{VHS}$ we assume to be the orientation-dependent field-induced rearrangement of the Dirac density of states due to Landau level formation, although renormalization of the Dirac energy dispersion such as indicated in ZrSiSe \cite{shao2020electronic} might also be important.


In conclusion, we present NMR results that delineate the effects of the symmetry-protected topological nodal line in ZrSiTe, and we also show that there is a van Hove singularity (VHS) close in energy which plays a significant role, and leads to a $T$-induced Lifshitz transition. DFT calculations confirm the presence of the VHS just below $E_F$. Sharp minima at 20 K in NMR shifts and $(T_1T)^{-1}$ are matched to a theoretical model indicating that the chemical potential crosses the protected nodal line along R-X. The results provide an indication of the 2D nature of this nodal line. Well-defined changes in the NMR results at higher temperatures indicate the influence of the 2D VHS, leading to a Lifshitz transition which occurs at 140 K with the field oriented perpendicular to the layers.

\begin{acknowledgments}
This work was supported by the Robert A. Welch Foundation, Grant No. A-1526 and Texas A\&M University. The financial support for sample preparation was provided by the National Science Foundation through the Penn State 2D Crystal Consortium-Materials Innovation Platform (2DCC-MIP) under National Science Foundation cooperative agreement DMR-1539916.
\end{acknowledgments}

%

\end{document}